# Design of a Non-vacuum-cooling Compact Scientific CCD Camera

Yi Feng, Hong-fei Zhang, Yi-ling Xu, Jin-ting Chen, Dong-xu Yang, Yi Zhang, Cheng Chen, Guang-yu Zhang, Jian-min Wang, Jian Wang, *Senior Member, IEEE*

*Abstract*—CCD was born in Bell Laboratories in 1969 and has been widely used in various fields. Its ultra-low noise and high quantum efficiency make it work well in particle physics, high energy physics, nuclear physics and astrophysics. Nowadays, more and more CCD cameras have been developed for medical diagnosis, scientific experiments, aerospace, military exploration and other fields. For the wide range of CCD cameras, a Non-vacuum-cooling compact (NVCC) scientific CCD camera has been developed, including FPGA-based low noise clock and bias driver circuit, data acquisition circuit, STM32-based temperature control design. At the same time, the readout noise of the imaging system is studied emphatically. The scheme to generate the CCD clock and the bias driving circuit through ultralow noise LDOs is proposed. The camera was tested in a variety of environments, and the test results show that the system can run at a maximum rate of 5M pixels/s and readout noise is as low as $9.29e^-$ when the CCD readout speed is 500K pixels/s. Finally, a series of stability tests were carried out on the camera system.

## I. INTRODUCTION

CCD was born in Bell Laboratories in 1969 and has been widely used in various fields. Its ultra-low noise and high quantum efficiency make it work well in particle physics, high energy physics, nuclear physics and astrophysics. Nowadays, CCD cameras are not only used in cutting-edge scientific research, but also play a very important role in industrial, commercial, medical and astronomical observation, such as industrial inspection, item scanning, medical diagnosis and planetary tracking.

We designed a NVCC scientific CCD camera which can be used in various fields, the overall structure shown in Figure 1.

The camera consists of CCD house, CCD controller and shutter. The CCD house contains a CCD chip (KAF-3200), a TEC (Thermo-Electric-Cooler) and a fan. In the camera, the CCD chip is cooled by the TEC, and the heat of the TEC is dissipated by the fan. At the same time, the CCD's temperature is monitored by the temperature sensor. The CCD controller is mainly composed of power module, clock and bias module, data acquisition module, FPGA module, USB interface module, temperature control module, shutter control module and fan control module. These modules cooperate with each other to implement the function of the CCD driving and the CCD output sampling. The shutter consists of stepper motors, blades and connector. Its minimum exposure time is as low as 0.2s.

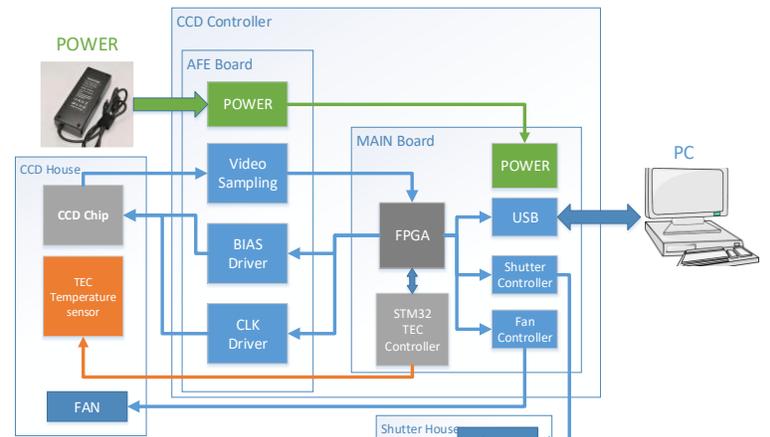

Figure 1 Structure diagram of the NVCC scientific CCD camera

## II. CONTROLLER DESIGN

The structure of the CCD controller is shown in Figure 2. It consists of two PCB boards: Front-end Circuit Board (FCB) and the Mother Board (MB), which are connected by an inter board connector. The FCB provides low noise power for the entire system, generates the clock and bias signals for the CCD, and samples the CCD signal. The MB controls the generation of clock and bias, receives the image data of the FCB, transfers data to the computer, and also controls the temperature through the STM32. In the design of the controller, the noise of the clock and bias module and the data acquisition module are deeply researched and analyzed. And we use various filtering processes to effectively reduce the readout noise of the system.

This work was supported by the National Natural Science Funds of China under Grant No: 11603023, 11773026, 11728509, the Fundamental Research Funds for the Central Universities (WK2360000003, WK2030040064), the Natural Science Funds of Anhui Province under Grant No: 1508085MA07, the Research Funds of the State Key Laboratory of Particle Detection and Electronics, the CAS Center for Excellence in Particle Physics, the Research Funds of Key Laboratory of Astronomical Optics & Technology, CAS.

The Authors Yi Feng, Hong-fei Zhang, Yi-ling Xu, Jin-ting Chen, Dong-xu Yang, Yi Zhang, Cheng Chen, Guang-yu Zhang, Jian-min Wang, Jian Wang is with the University of Science and Technology of China, Jian Wang, State Key Laboratory of Technologies of Particle Detection and Electronics, University of Science and Technology of China, Hefei, Anhui 230026, China (e-mail: Hong-fei Zhang, nghong@ustc.edu.cn; Jian Wang, wangjian@ustc.edu.cn).

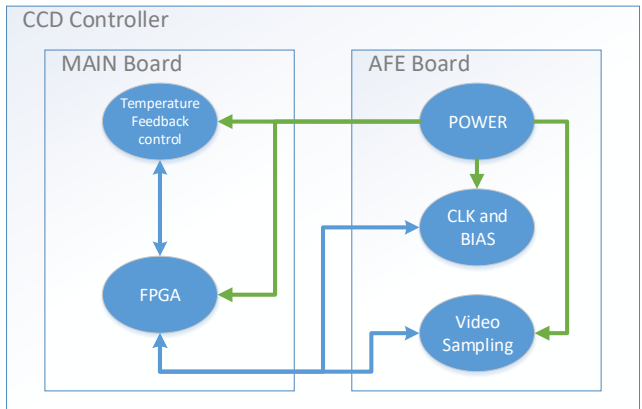

Figure 2 controller of the NVCC CCD camera

Now, we have completed the entire system design, and conducted a comprehensive test of camera performance, including safety, stability, reliability. The NVCC CCD camera is shown in Figure 3.

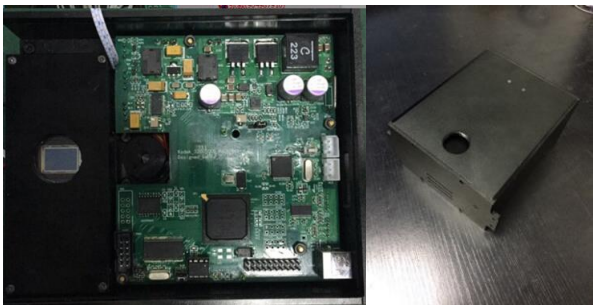

Figure 3 Pictures of the NVCC CCD camera

### III. SYSTEM TEST

What's more, we also tested the noise and gain of the camera. The results are shown in Table 1.

TABLE I
NOISE AND GAIN OF THE CAMERA

| Readout Rate(pixels/s) | Readout Noise($e^-$) Low Gain($e^-/ADU$) | Readout Noise($e^-$) High Gain($e^-/ADU$) |
|---|---|---|
| 500K | 9.29 0.53 | 18.88 1.11 |
| 1M | 10.11 0.56 | 16.73 1.09 |
| 2.5M | 11.47 0.59 | 16.04 1.08 |
| 5M | -- | 17.18 1.05 |

According to the test results, the camera can read data at a maximum rate of 5M pixels/s and readout noise is as low as $9.29e^-$ at the speed of 500K pixels/s. Finally, we also tested the dark current of the camera which is $0.0147e^-/s$.

Currently, one of the camera's uses is for De-Lingha Observatory. The telescope is shown in Figure 4.

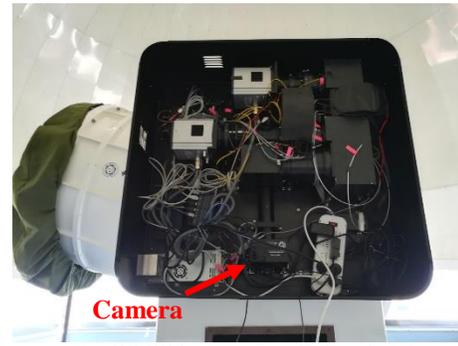

Figure 4 NVCC CCD camera installed on the telescope

We use the camera for guiding. The image is shown in Figure 5.

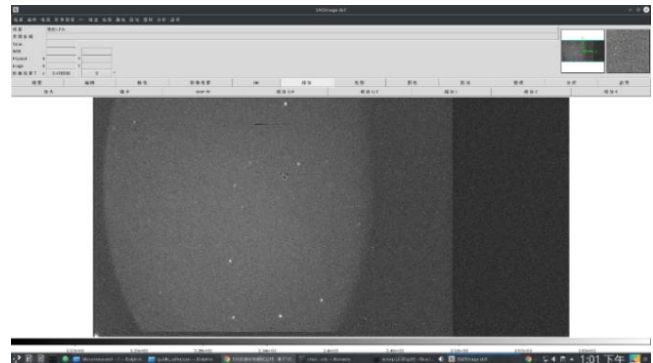

Figure 5 Image taken from guiding system

In addition, we have conducted tests in a broad ambient temperature range on the camera for a long period of time, and the CCD can operate at a temperature 45 ° C lower than the ambient temperature without performance degradation, verifying the stability of the camera.

### IV. CONCLUSION

After testing, the NVCC CCD camera has stable performance. The readout noise is as low as $9.29e^-$ at the speed of 500K pixels/s, and the dark current is as low as $0.0147e^-/s$. It is currently applied to the De-Lingha Observatory for astronomical observations. After a long time of running, this is a stable, high-performance scientific CCD camera.

REFERENCE

[1]. Yuan, X. Y., Cui, X. Q., Liu, G. R., et al, "Chinese Small Telescope Array (CSTAR) for Antarctic Dome A", SPIE Proc. 7012, 70124G (2008).
[2]. Xu Zhou, Zhen-Yu Wu, Zhao-Ji Jiang, et al. "Testing and data reduction of the Chinese Small Telescope Array (CSTAR) for Dome A, Antarctica", Research in Astron. Astrophys , 10(3):279-290 (2010).
[3]. Cui, X. Q., Yuan, X. Y., Gong, X. F., "Antarctic Schmidt Telescopes (AST3) for Dome A", Proc. SPIE Proc, 7012, 70122D (2008).
[4]. Zhang H F, Wang J M, Tang Q J, et al. Design of Ultra-Low Noise and Low Temperature Usable Power System for High-Precision Detectors [J]. IEEE Transactions on Nuclear Science, 63(6): 2757-2763 (2016).
[5]. Jiang W, Jiang X. Design of an intelligent temperature control system based on the fuzzy self-tuning PID [J]. Procedia Engineering, 43: 307-311 (2012).